\documentclass[12pt]{iopart}
\usepackage{iopams} 
\usepackage{graphicx}
\usepackage{subfigure}
\usepackage{placeins}
\usepackage{url}
\usepackage{multirow}
\usepackage{cleveref}
\begin{document}
	
\title{Vibration Analysis of KAGRA Cryostat at Cryogenic Temperature}
\author{R.Bajpai$^1$, T.Tomaru$^2$, N.Kimura$^3$, T.Ushiba$^3$, K.Yamamoto$^4$, T.Suzuki$^5$, T. Honda$^6$}

\address{$^1$ The Graduate University for Advanced Studies, Department of Accelerator Science, School of High Energy Accelerator Science, High Energy Accelerator Research Organization (KEK) Tsukuba, Ibaraki, 305-0801, Japan}

\address{$^2$ National Astronomical Observatory of Japan, 2 Chome-21-1 Osawa, Mitaka, Tokyo, 181-8588, Japan}

\address{$^3$ The University of Tokyo Institute for Cosmic Ray Research Kamioka Observatory, Higashimozumi 238, Kamioka, Hida, Gifu, 506-1205, Japan}

\address{$^4$ University of Toyama, 3190 Gofuku, Toyama, 930-8555, Japan}

\address{$^5$ The University of Tokyo Institute for Cosmic Ray Research, Kashiwanoha 5-1-5, Kashiwa, Chiba, 277-8582, Japan}

\address{$^6$ High Energy Accelerator Research Organisation, 1-1 Oho, Tsukuba, Ibaraki, 305-0801, Japan}

\ead{bajpai@post.kek.jp} 

\begin{abstract}
KAGRA uses cryogenics to cool its sapphire test masses down to 20 K to reduce the thermal noise. 
However, cryocooler vibration and structural resonances of the cryostat couple to test mass and can contaminate the detector sensitivity. We performed vibration analysis of the cooling system at cryogenic temperature to study its impact on detector sensitivity. 
Our measurement show shield vibration below 1 Hz is not impacted by cryocooler operation or structural resonances and follows ground motion. 
The noise floor of the shield in 1-100 Hz was observed to be 2-3 order of magnitude larger than seismic motion even without cryocooler operation. 
The operation of cryocoolers does not change the noise floor, but 2.0 Hz peaks and their harmonics were observed over the entire spectrum (1-100 Hz). 
These results were used to calculate the coupling of cooling system vibration to the test mass.
We conclude that vibration from the cooling system does not limit KAGRA design sensitivity.
\end{abstract}



\section{Introduction}

On 14 September 2015, LIGO Hanford and Livingston made the first direct detection of gravitational wave (GW150914) \cite{1}, since then several detections \cite{2,3} have been reported by Advanced LIGO \cite{4} and Advanced Virgo \cite{5}, accelerating the field of GW astronomy. 
With these second-generation detectors expected to achieve peak sensitivity, studies are already underway to develop third-generation detectors like Einstein Telescope \cite{6} and Cosmic Explorer \cite{7}. 
Einstein Telescope will be constructed underground (to reduce seismic noise) and will employ cryogenics to cool down the mirrors (to reduce thermal noise), technologies employed in KAGRA; making it an excellent case study for future gravitational wave detectors.

Large-scale Cryogenic Gravitational wave Telescope (KAGRA) \cite{8} is a Fabry-Perot Michelson Interferometer based GW detector located in Kamioka mine, Japan. 
KAGRA is under commissioning and will join Advanced LIGO and Advanced Virgo for the next observation run. 
In low-frequency, fundamental noise sources limiting the sensitivity of interferometer based detectors are the seismic, Newtonian and thermal noise. 
While the fundamental design of KAGRA is similar to other second-generation detectors, the underground location and cryogenic operation of the four main mirrors are two features unique to KAGRA. 
The underground location provides a quiet site with low seismic and gravity gradient noise \cite{9}, while the cryogenic operation cools the mirrors down to 20 K, reducing the thermal noise. 
Vibration originating from the cryocooler operation and structural resonances of the cryostat can contaminate the detector sensitivity making the quiet underground location redundant. 
Therefore, monitoring and characterization of the vibration inside cryostat is critical for the optimum noise performance of KAGRA.

In April 2020, KAGRA conducted an international observation run, "O3GK"\cite{10}, along with GEO600 \cite{11}. 
Several noise sources were identified during this run, and a noise budget was prepared \cite{10}. 
However, as the mirrors were not cooled, the noise contribution from the cooling system was estimated based on room temperature, in vacuum vibration measurement performed 2.5 years before O3GK. 
During the upcoming observation run, the mirrors will be cooled down, so we performed vibration analysis of one of the KAGRA cryostats at 12 K. 
In this paper, we first describe KAGRA cooling system then report the results of vibration analysis at cryogenic temperature.

\section{Experiment Layout} \label{sec:2}

\subsection{KAGRA Cooling System}
\begin{figure}[bp!]
	\centering
	\subfigure[Schematic layout of KAGRA cryogenic system. Platform (PF), marionette (MN), intermediate mass (IM), test mass (TM) and their recoil masses MNR, IRM, RM form the "Cryogenic Payload", cooled inside a double radiation shield cryostat. The outer black block is the vacuum chamber; the inner boxes are 80 K and 8 K radiation shield; 80 K shield is cooled down by 1st stage of PTC-1,2,3 and 4 while 8 K shield is cooled by 2nd stage of PTC-2 and 4. The 2nd stage of PTC-1 and 3 are connected to the cooling bar and cool down the payload through soft 6N Al heat links. A pair of single-stage PTC cools the duct shield on either side of the cryostat down to 120 K.  The pink arrows represent the vibration transfer path. This path is: top of inner shield $\rightarrow$ HLVIS $\rightarrow$ HL  $\rightarrow$ MNR $\rightarrow$ PF $\rightarrow$ MN $\rightarrow$ IM $\rightarrow$ TM.] {\includegraphics[width=0.9\textwidth]{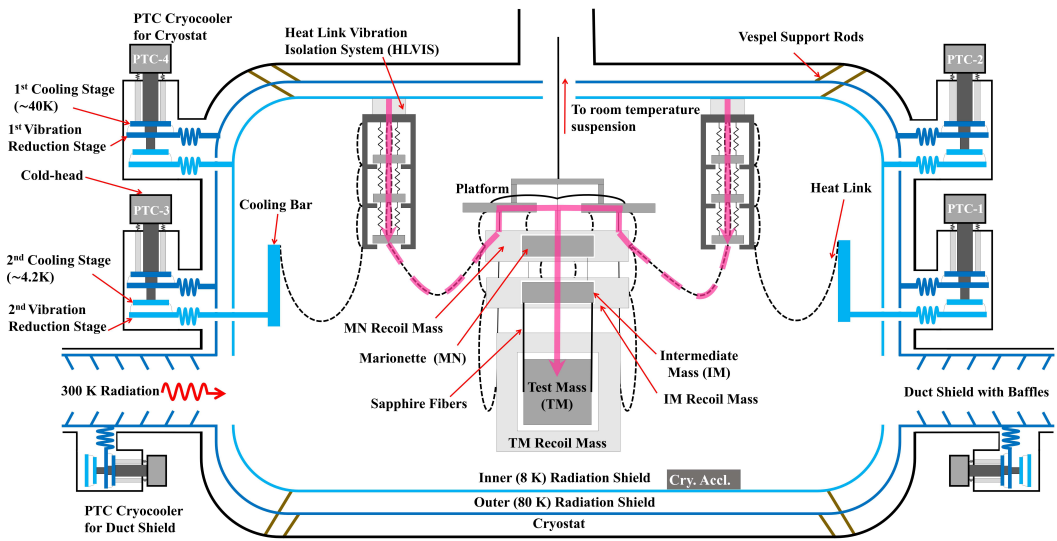}\label{fig:1-A}}	
	\subfigure[Technical illustration of KAGRA Cryostat. The cryostat chamber (Height: 4.3 m and Outer Diameter: 2.6 m) is made of Stainless Steel (SUS-304) and weighs about 11000 kg. The 80 and 8 K shields are made of Aluminium (Al1070) and have a combined weight of 1400 kg. Each shield is supported by 8 rods made of DuPont Vespel$\textregistered$SP-1, with other ends of the rod fixed to the cryostat chamber.] {\includegraphics[width=0.9\textwidth]{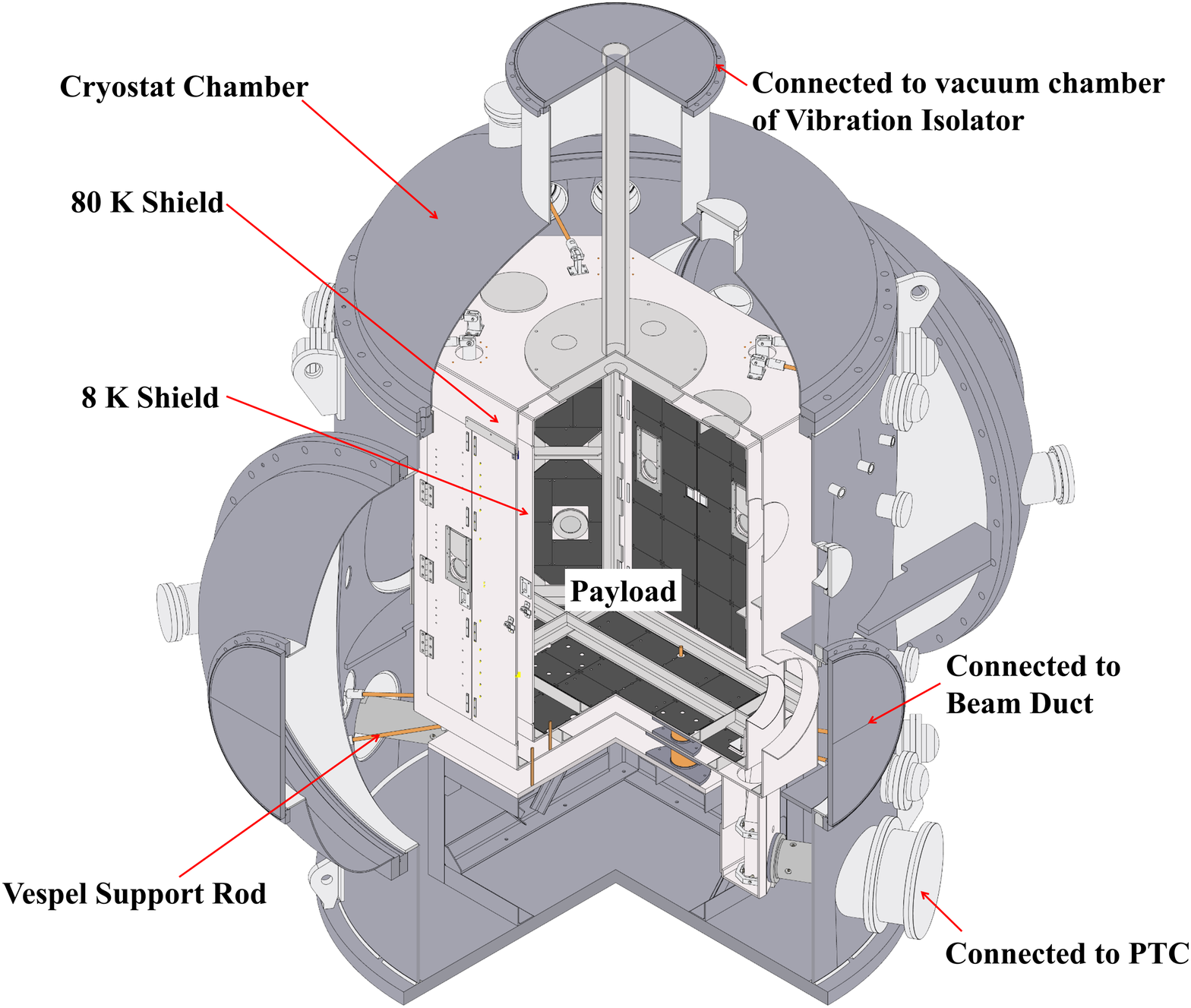}\label{fig:1-B}} 
	\caption{Schematic layout of KAGRA cooling system and technical illustration of KAGRA cryostat.}
\end{figure}
	Mirror (23 kg sapphire test mass) used for Fabry-Perot cavity in KAGRA are suspended from a nine-stage suspension system called 'Type-A Suspension' to isolate it from seismic motion. 
	The top five stages collectively, called 'Type-A tower', are kept at room temperature, while the bottom four stages are called 'Cryogenic Payload' \cite{12}. 
	It is cooled within a double-radiation shield cryostat using four ultra-low vibration pulse-tube cryocoolers (PTC)\cite{13}. 
	This PTC has two cold stages (cooled to $\sim 40$ K and $\sim 4$ K) in thermal contact with their Vibration Reduction (VR) stage. 
	The 1st VR stage of each cryocooler is connected to the outer (80 K) shield, while 2nd stage of two cryocoolers is connected to the inner (8 K) shield. 
	Thermal radiation to shields effectively cools the payload down to $\sim 100$ K. 
	2nd stage of the other two cryocoolers are connected to 6N (6 Nine, $99.9999\%$ purity) aluminum  cooling bar (rigidly bolted to 8 K shield but thermally isolated) that are in thermal contact with the payload through thin 6N Al heat-links \cite{14} to facilitate conduction cooling of the test mass down to 20 K.
	Each shield is supported by 8 Vespel SP-1 rods (4 at top, 4 at bottom) with other ends fixed to the cryostat chamber. These rods are designed to be rigid at cryogenic temperature, considering their thermal shrink.
	
 Another source for heat radiation are the holes on either side of the shields that make the optical path for the laser. 
 To minimize the 300 K radiation from these holes, 5-m long cold pipes, called duct shields \cite{15}, are installed on either side. 
 Each duct shield is cooled down by a double-stage PTC \cite{13}. Towards next observation runs, we have already replaced the single-stage PTCs for duct-shield cooling to double-stage PTCs. \Cref{fig:1-A,fig:1-B} show the KAGRA cooling system's layout and technical illustration of the cryostat, respectively.

\subsection{Setup}\label{sec:2.2}
\begin{figure} [h!]
 	\centering
 	\includegraphics[width=1\textwidth]{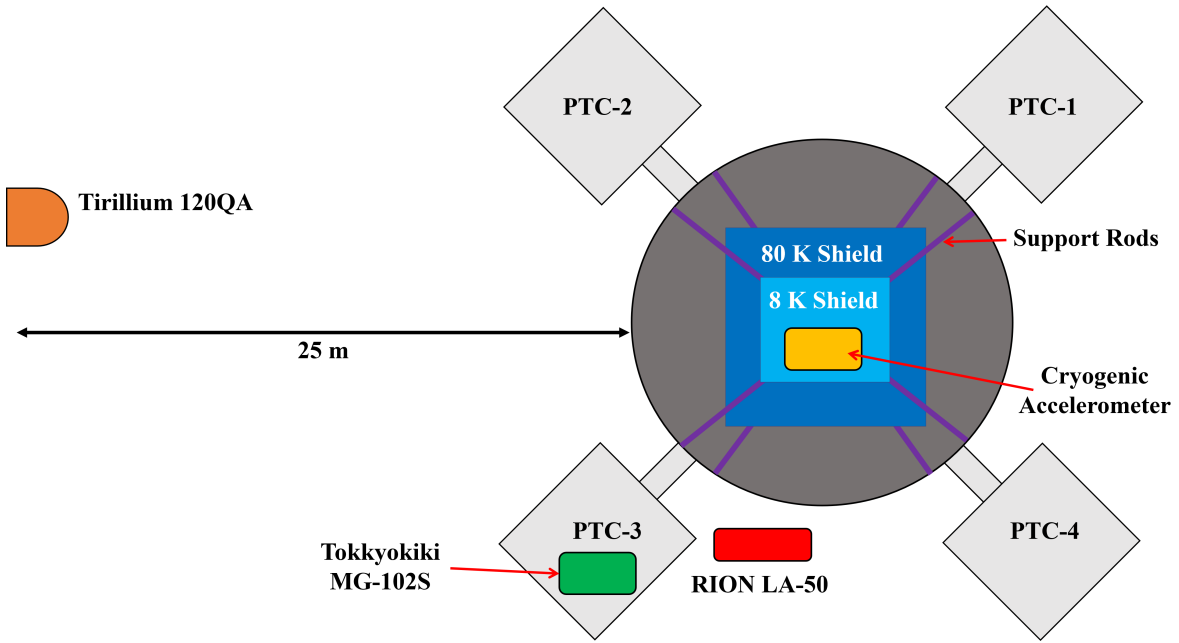}
 	\caption{Setup for vibration analysis experiment using the IY Cryostat of KAGRA showing the location of the 4  accelerometers.}
 	\label{fig:2}
 \end{figure}

 Cryocooler vibration and structural resonances of the shield can couple to test mass through the heat links. 
 The vibration coupling path is: top of inner shield $\rightarrow$ HLVIS $\rightarrow$ HL  $\rightarrow$ MNR $\rightarrow$ PF $\rightarrow$ MN $\rightarrow$ IM $\rightarrow$ TM, as denoted by pink arrows in \cref{fig:1-A}.
 To evaluate its impact on detector sensitivity, we monitored inner shield vibration using a cryogenic accelerometer \cite{16} developed specifically for this purpose. 
 We used IY Cryostat \cite{8} of KAGRA for this measurement. 
 A reference accelerometer, \textit{RION LA-50} placed next to the cryostat and a reference seismometer, \textit{TRILLIUM 120QA}, placed 25 m from the cryostat were used to monitor the seismic motion. 
 Another reference accelerometer, \textit{TOKKYOKIKI MG-102S}, mounted on PTC-3, monitored cryocooler vibration. 
 \Cref{fig:2} shows the position where each accelerometer was located. 
 
\section{Measurement Results}

 We performed a series of coincident measurements at several temperatures down to 12 K and characterized the vibration of the cryostat. 
 Radiation shield vibration was measured in 0.1-100 Hz bandwidth because vibration above 100 Hz is well attenuated by the heat-links and cryogenic payload. 
 In this section, the result of these measurements is reported.
\subsection{Influence of Ground Motion}

\begin{figure} [h!]
	\centering
	\includegraphics[width=1.1\textwidth]{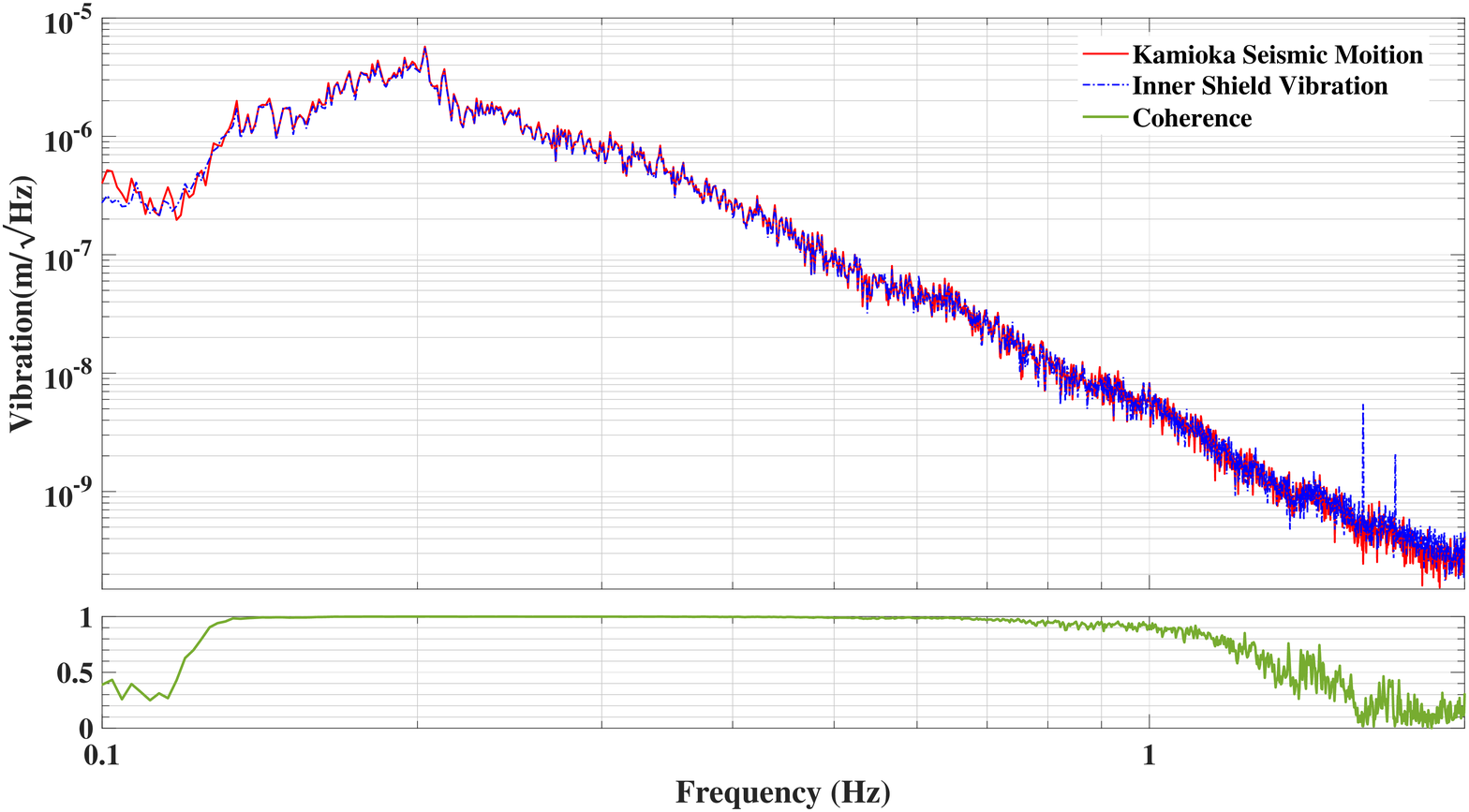}
	\caption{Comparison of vibration spectrum of inner radiation shield at 12 K(blue) and Kamioka seismic motion (red), in 0.1-2 Hz band showing that the shield follows ground motion below 1 Hz. The green curve shows coherence between the two.}
	\label{fig:3}
\end{figure}

\Cref{fig:3} shows radiation shield vibration at 12 K (blue) (measured using cryogenic accelerometer), KAGRA seismic motion (red) (measured using \textit{TRILLIUM 120QA}) and coherence (green) between the two spectra in 0.1-2 Hz band. 
Below 1 Hz, the two spectra widely coincide, and coherence between the two is almost 1. 
Above 1 Hz motion of the shield starts to deviate from the seismic motion; this is due to internal resonances and cryocooler operation. 
We conclude that the shield follows ground motion below 1 Hz and focus on vibration spectrum in 1-100 Hz bandwidth for further measurements. 

\subsection{Influence of Shield and Cryostat Resonances} \label{sec:3.2}
\begin{figure} [t]
	\centering
	\includegraphics[width=1.1\textwidth]{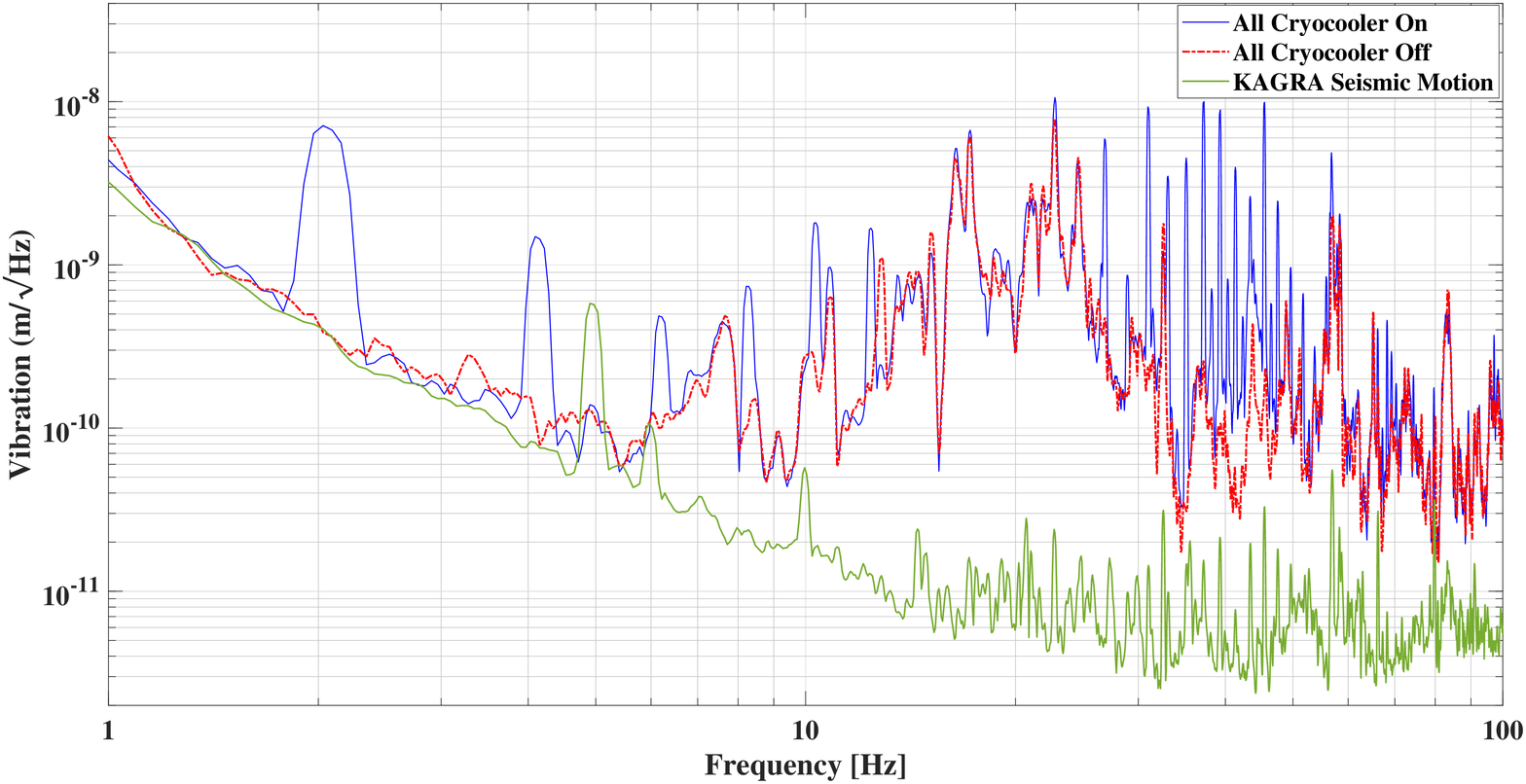}
	\caption{Comparison of radiation shield vibration spectra at 12 K with cryocooler turned on (blue) and off (red). The green spectrum is seismic motion at KAGRA site.}
	\label{fig:4}
\end{figure}

We compare the vibration spectra of the radiation shield when cryocoolers are turned on (blue) and off (red) at 12 K in \cref{fig:4}. 
When the cryocoolers are turned off, the peaks that appear in the radiation shield vibration spectra are due to internal resonances of chamber and shield as they are not affected by cryocooler operation. 
We conclude that cryocooler operation does not affect the noise floor (background vibration due to internal resonances), but 2.0 Hz peaks and its harmonics appear over the entire spectra.

To distinguish between structural resonances (red spectra in \cref{fig:4}) from radiation shield and chamber, we performed a hammering test and concurrent measurement of shield and chamber vibration, respectively, at room temperature.
A \textit{TOKKYOKIKI MG-102S} accelerometer was mounted on the bottom of the radiation shield and the shield was excited using an impact hammer. 
The peaks observed in the frequency response of the accelerometer were concluded to be shield resonances.

To identify chamber resonances, the vibration at top of the cryostat chamber and bottom of inner shield were simultaneously measured using \textit{TOKKYOKIKI MG-102S} and \textit{RION LA-50}, respectively. 
The peaks with coherence $>$ 0.9 were concluded to be chamber structural resonances. 
Note that during this simultaneous measurement no external excitation was introduced.
The origin of resonance peaks identified are summarized in \cref{tab:1}.
\begin{table}[hbt!]
	\caption{Resonant frequency identified by hammering test for radiation shield and chamber.}
	\label{tab:1}
	\begin{center}
		\begin{tabular}{|l|p{0.75\textwidth}|}
			\hline
			\textbf{Origin} & \textbf{Frequency (Hz)}  \\
			\hline
			Chamber & 9.1, 12.4, 17.2, 21.8, 22, 28.2, 28.5, 34.2, 39, 39.5, 48.5, 48.9, 49.2, 49.7, 50.6, 51.2, 52, 52.2, 52.7, 53.3, 53.62, 55.1, 55.7, 77.3, 83.7\\
			\hline
			Radiation Shield & 16.3, 23.7, 31.3, 37.2, 42.5,47.1, 50.3, 54.1, 57.8, 64.4, 68.5, 73.4, 76.1, 78.8, 83.4, 90.4  \\
			\hline
		\end{tabular}
	\end{center}
\end{table}
 
Comparing shield vibration with  KAGRA seismic motion (green spectra in  \cref{fig:4}), measured using \textit{RION LA-50} we observed that vibration of the shield is 2-3 order of magnitude larger than the seismic motion. 
The region around 10-25 Hz, is dominated by internal resonances of the cooling system whereas, above 25 Hz, several peaks with a magnitude of $10^{-8}$ m/$\sqrt{\mathrm{Hz}}$ exist due to cryocooler operation.

\subsection{Influence of Cryocooler Operation}\label{sec:3.3}

\begin{figure} [h!]
	\centering
	\includegraphics[width=1.1\textwidth]{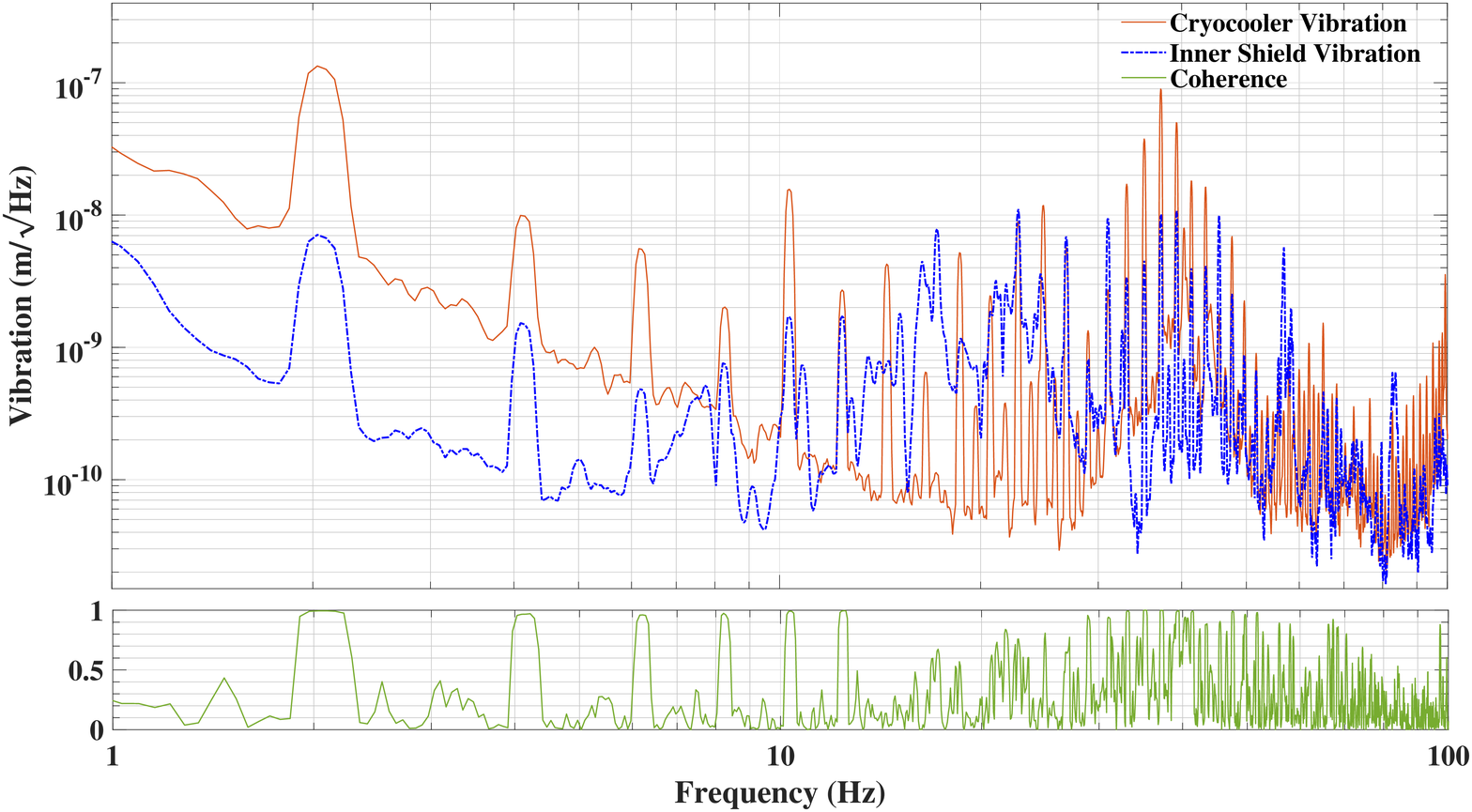}
	\caption{Comparison of vibration spectrum of inner radiation shield at 12 K (blue) and cryocooler cold-head vibration (orange) in 1-100 Hz band showing that cryocooler operation introduces 2.0 Hz and it's harmonics over the entire spectrum. Green curve shows coherence between the two.}
	\label{fig:5}
\end{figure}

\Cref{fig:5} shows the radiation shield vibration at 12 K(blue), cryocooler vibration (orange) and  coherence between the two spectrums in 1-100 Hz band while all cryocoolers are operating. 
Note that the reference accelerometer mounted on PTC-3 (see \fref{fig:2}) monitors the vibration of cryocooler cold-head \cite{13} (see \cref{fig:1-A}) and not that of the vibration reduction stages. 
We observed that peaks of 2.0 Hz and its harmonics appear over entire radiation shield vibration spectrum; these peaks are an order lower in magnitude with coherence close to 1 in comparison with cold-head vibration. 
The radiation shield spectrum is dominated by cryostat internal resonances in 10-30 Hz band which have a much larger magnitude compared to cryocooler vibration coupling, as discussed in \cref{sec:3.2}. 

\section{Vibration Coupling through Heat Links}
\begin{figure} [h!]
	\centering
	\includegraphics[width=1.1\textwidth]{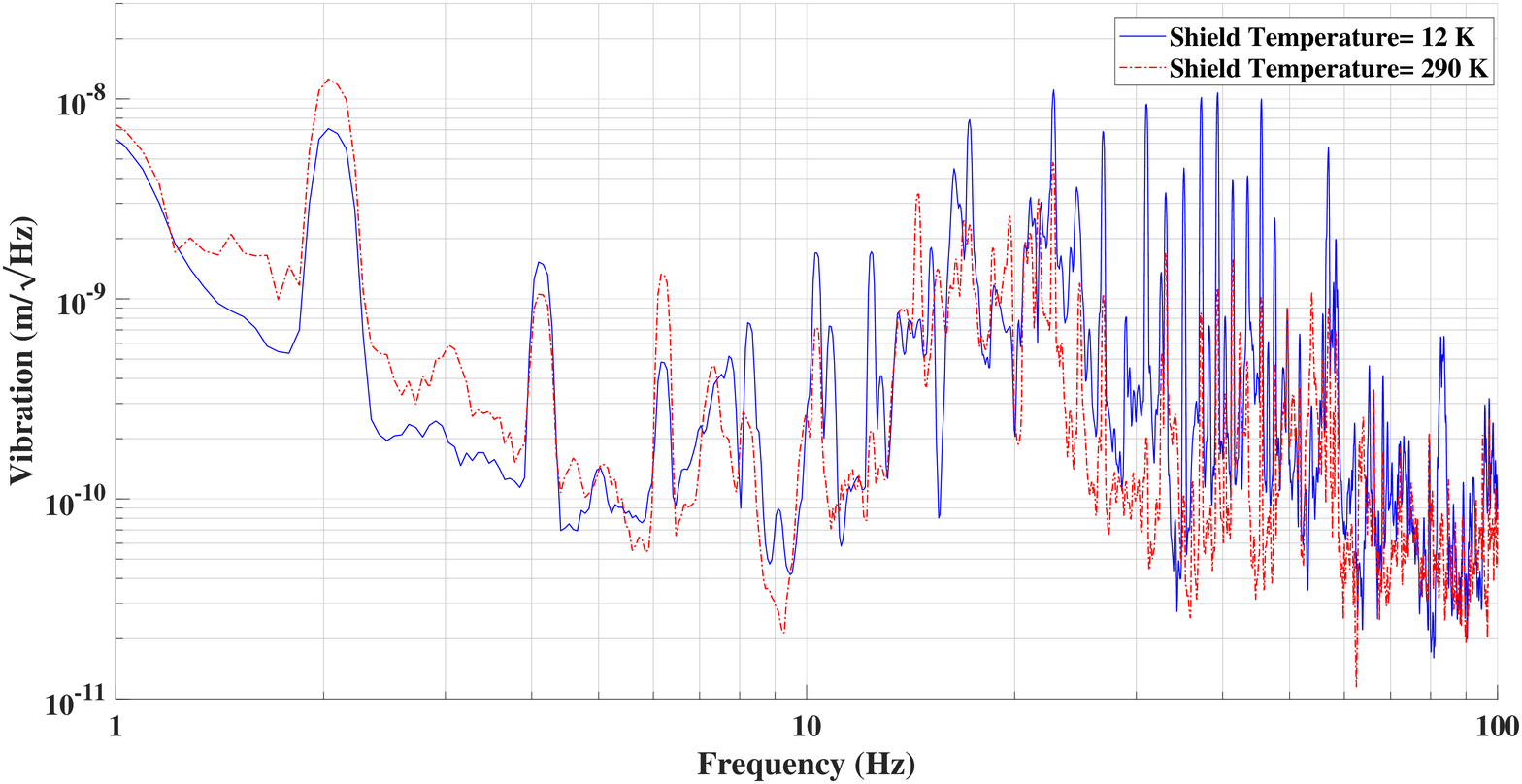}
	\caption{Comparison of radiation of shield vibration at 290 K (red) and 12 K (blue), showing increase in resonance peaks and coupling of cryocooler vibration.}
	\label{fig:6}
\end{figure}

One of the primary motivation for this study was to evaluate whether shield vibration at cryogenic temperature will contaminate the detector sensitivity because the current estimate is based on in vacuum, room temperature measurement.
\Cref{fig:6} shows the comparison of radiation shield vibration spectra at 290 K (red) and 12 K (blue).
We observed that there was significant increase in magnitude of resonance peaks and cryocooler vibration  coupling.
Considering this increased vibration magnitude at cryogenic temperature we re-evaluated the vibration coupling through heat-links.
In this section, we show the details and result of vibration coupling calculations.
\subsection{Vibration Coupling}
The vibration at top of the radiation shield couples to the test mass as denoted by the pink arrows in \cref{fig:1-A}. 
Now, both horizontal and vertical vibration couples to the TM, based on the vibration path test mass displacement are derived below.
\begin{itemize}
	\item \textit{Horizontal vibration coupling:} 
	Test-Mass displacement due to horizontal vibration will be:
	\begin{equation} \label{1}
	{\mathrm{Displacement}^{\mathrm{H}}_{\mathrm{TM}}}=V^{\mathrm{H}}_{\mathrm{Top}}\times TF^{\mathrm{H}}_{\mathrm{HLVIS}}\times TF^{\mathrm{H}}_{\mathrm{HL}}\times TF^{\mathrm{H}}_{\mathrm{MNR \rightarrow TM}}
	\end{equation}
	where $V^{\mathrm{H}}_{\mathrm{Top}}$ is the horizontal vibration at top of the inner shield and, $TF^{\mathrm{H}}_{\mathrm{component}}$ is the horizontal-transfer function of the component.
	Now, the vibration at the top ($V^{\mathrm{H}}_{\mathrm{Top}}$) can be derived from the measured vibration of the bottom ($V^{\mathrm{H}}_{\mathrm{Bottom}}$) as: 
	\begin{equation} \label{2}
	V^{\mathrm{H}}_{\mathrm{Top}}= V^{\mathrm{H}}_{\mathrm{Bottom}} \times TF^{\mathrm{H}}_{\mathrm{Bottom \rightarrow Top}}  
	\end{equation}
	where $V^{\mathrm{H}}_{\mathrm{Bottom}}$ is the blue spectrum in \cref{fig:6} and $TF^{\mathrm{H}}_\mathrm{{Bottom\rightarrow Top}}$ is vibration transfer from bottom to top of the shield along horizontal direction.
	
	\item \textit{Vertical vibration coupling:} 
	As KAGRA tunnel has $\approx$ 1/300 inclination for water drainage, gravity direction isn't perpendicular to the optical axis and we assume 1\% vertical to horizontal vibration coupling (commonly used, considering the slope and imperfections in the vibration isolation system).
	The displacement of TM along horizontal axis due to vertical vibration coupling will be:
	\begin{equation} \label{3}
	{\mathrm{Displacement}^{\mathrm{V}}_{\mathrm{TM}}}=V^{\mathrm{V}}_{\mathrm{Top}}\times TF^{\mathrm{V}}_{\mathrm{HLVIS}}\times TF^{\mathrm{V}}_{\mathrm{HL}}\times TF^{\mathrm{V}}_{\mathrm{MNR \rightarrow TM}} \times 1\%
	\end{equation}
	where $V^{\mathrm{V}}_{\mathrm{Top}}$ is the vertical vibration at top of the inner shield and, $TF^{\mathrm{V}}_{\mathrm{component}}$ is the vertical-transfer function of the component.
	Furthermore, vertical and horizontal vibration at top of the radiation shield are assumed to be same $(V^{\mathrm{V}}_{\mathrm{Top}}=V^{\mathrm{H}}_{\mathrm{Top}})$. 
\end{itemize}
  Note that the vibration coupling/test-mass displacement calculated in \cref{1,3} is due to a single heat-link (between HLVIS and MNR). The details of how each transfer function ($TF^{\mathrm{H}}_{\mathrm{component}}$, $TF^{\mathrm{V}}_{\mathrm{component}}$) in \cref{1,3} was evaluated is described below:
  \begin{itemize}
  	\item $TF^{\mathrm{H}}_\mathrm{{Bottom\rightarrow Top}}:$ 
  	The vibration transfer from bottom to top of the shield was assumed to be same between room and cryogenic temperature because density and elastic constant of Vespel support rods do not vary largely with temperature. 
  	As the transfer function couldn't be measured practically owing to the size of the shield, we simply measured the ratio of vibration at the top and bottom of the shield at 300 K (measured using \textit{TOKKYOKIKI MG-102S} accelerometer), and substituted it as $TF_\mathrm{{bottom\rightarrow top}}$.
  	
  	\item $TF_{\mathrm{HLVIS}}:$ 
  	The heat link vibration isolation system (HLVIS) transfer function was practically evaluated at 16 K. 
  	The details of HLVIS, its performance evaluation and transfer function can be found in \cite{17,18}.
  	
  	\item $TF_{\mathrm{Heat-Link}}:$ 
  	Heat-link transfer function was evaluated by measuring its spring constant at 300 K. 
  	This transfer function and details of its measurement can be found in \cite{17,18}.
  	
  	\item $TF_{\mathrm{MNR\rightarrow TM}}:$ 
  	As shown in \cref{sec:2.2} vibration transfers from MNR to the TM. This mechanical transfer function from MNR to TM was calculated with rigid body modelling tool, SUMCOM \cite{19}, developed for KAGRA suspension characterization.
  	
  \end{itemize} 
  
  \subsection{Impact on Sensitivity}
\begin{figure} [h]
	\centering
	\includegraphics[width=1.1\textwidth]{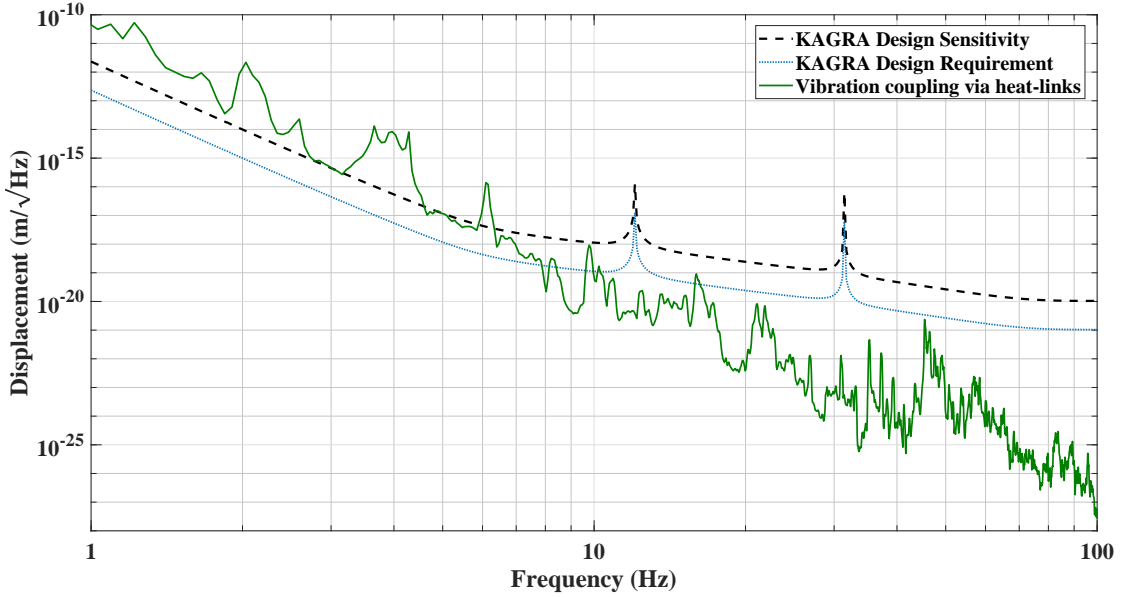}
	\caption{Radiation shield vibration coupling to TM compared with KAGRA design sensitivity and requirement. The details of KAGRA design sensitivity can be found at \cite{8}. The sensitivity requirement is set as one-tenth of the design sensitivity to achieve 10 SNR.}
	\label{fig:7}
\end{figure}

  Substituting the transfer functions and measured vibration in \cref{1,2,3}  we get the displacement of TM due to vertical and horizontal vibration coupling for a single heat-link.
  As there are 4-heat-links connected between HLVIS and MNR a factor of 4 is multiplied. Furthermore, similar vibration coupling with no correlation for all four test masses is expected, so a factor of $\sqrt{4}$ is also multiplied. Therefore, the total displacement of a test-mass due to radiation shield vibration becomes:
  \begin{equation}\label{4}
  {\mathrm{Displacement}_{\mathrm{TM}}}=4 \times \sqrt {4} \times\left[ {\mathrm{Displacement}^{\mathrm{V}}_{\mathrm{TM}}}^2+{\mathrm{Displacement}^{\mathrm{H}}_{\mathrm{TM}}}^2 \right]^{1/2}
  \end{equation} 
  Note that in \cref{4}, displacement due to vertical and horizontal vibration are added in quadrature as they will be incoherent.
  The test mass displacement due vertical and horizontal vibration coupling to the test mass and KAGRA design sensitivity is plotted in \cref{fig:7}. 
   From this, we expect that vibration coupling through the heat-links will not be an issue for KAGRA as it is below the design requirement in observation band over dozens hertz.

\FloatBarrier
\section{Summary}
The test masses in KAGRA are cooled down to reduce the thermal noise. 
However, the heat-links used to cool down the mirror couple the radiation shield vibration to test mass. 
The current noise budget for vibration coupling through the heat-links is based on in vacuum, room temperature measurement. 
With the goal to update this noise budget, we performed vibration analysis of one of the KAGRA cryostat at 12 K and calculated if this vibration impacts the detector sensitivity. 
Our study shows that the vibration of the radiation shield is 2-3 order of magnitude larger than Kamioka seismic motion in 1-100 Hz band due to internal resonance of the cooling system and operation of the cryocooler. 
Based on the vibration analysis results, the vibration coupling via heat-links was calculated, and it was found to be below the KAGRA design requirement and should not be an issue.

\section*{Data Availability}
The data that support the findings of this study are available from the corresponding author upon reasonable request.

\section*{Acknowledgment}
This work was supported by MEXT, JSPS Leading-edge Research Infrastructure Program, JSPS Grant-in-Aid for Specially Promoted Research 26000005, JSPS Grant-in-Aid for Scientific Research on Innovative Areas 2905: JP17H06358, JP17H06361 and JP17H06364, JSPS Core-to-Core Program A. Advanced Research Networks, JSPS Grant-in-Aid for Scientific Research (S) 17H06133, the joint research program of the Institute for Cosmic Ray Research, University of Tokyo, National Research Foundation (NRF) and Computing Infrastructure Project of KISTI-GSDC in Korea, Academia Sinica (AS), AS Grid Center (ASGC) and the Ministry of Science and Technology (MoST) in Taiwan under grants including AS-CDA-105-M06, the LIGO project, and the Virgo project. We would also like to thank PEM group for providing impact hammer for the hammering test and DGS group for letting us use KAGRA data acquisition system for some of the measurements presented in this paper.

\end{document}